\documentclass[%
 reprint,
 amsmath,amssymb,
prl,
]{revtex4-1}

\usepackage{tikz}
\usetikzlibrary{"arrows", "automata", "backgrounds", "calendar", "chains", "matrix", "mindmap", "patterns", "petri", "shadows", "shapes.geometric", "shapes.misc", "spy", "trees"}
\usetikzlibrary{arrows,shapes}
\usetikzlibrary{trees}
\usetikzlibrary{matrix,arrows} 	
\usetikzlibrary{positioning}				
\usetikzlibrary{calc,through}				
\usetikzlibrary{decorations.pathreplacing}
\usepackage{pgffor}

\usetikzlibrary{decorations.pathmorphing}	
\usetikzlibrary{decorations.markings}
\tikzset{
    vector/.style={decorate, decoration={snake}, draw},
	provector/.style={decorate, decoration={snake,amplitude=2.5pt}, draw},
	antivector/.style={decorate, decoration={snake,amplitude=-2.5pt}, draw},
    fermion/.style={draw=black, postaction={decorate},
        decoration={markings,mark=at position .55 with {\arrow[draw=black]{>}}}},
    fermionbar/.style={draw=black, postaction={decorate},
        decoration={markings,mark=at position .55 with {\arrow[draw=black]{<}}}},
    fermionnoarrow/.style={draw=black},
    sigma/.style={draw=green},    
    gluon/.style={decorate, draw=black,
        decoration={coil,amplitude=1.5pt, segment length=3pt}},
    phi/.style={dashed, draw=blue},
    scalar/.style={dashed,draw=black, postaction={decorate}},
    scalarbar/.style={dashed,draw=black, postaction={decorate},
        decoration={markings,mark=at position .55 with {\arrow[draw=black]{<}}}},
    scalarnoarrow/.style={dashed,draw=black},
    electron/.style={draw=black, postaction={decorate},
        decoration={markings,mark=at position .35 with {\arrow[draw=black]{>}}}},
    positron/.style={draw=black, postaction={decorate},
        decoration={markings,mark=at position .35 with {\arrow[draw=black]{<}}}},        
    bigvector/.style={decorate, decoration={snake,amplitude=4pt}, draw},
}

\usepackage{graphicx,color}
\usepackage{latexsym,amsmath,amssymb,graphicx,booktabs}
\usepackage{hyperref}
\usepackage{stackrel}

\usepackage[english]{babel}
\usepackage{amsmath}
\usepackage{amssymb}
\usepackage{amsbsy}
\usepackage{amstext}
\usepackage{bm}
\usepackage{graphicx}
\usepackage{subfigure}
\usepackage{multirow,rotating}
\usepackage{verbatim}

\graphicspath{{./figures/}}

\begin{document}

\newcommand{\be}{\begin{eqnarray}}
\newcommand{\ee}{\end{eqnarray}}
\newcommand{\bdm}{\begin{displaymath}}
\newcommand{\edm}{\end{displaymath}}
\newcommand{\ds}{\displaystyle}
\newcommand{\ba}{\begin{array}}
\newcommand{\ea}{\end{array}}
\newcommand{\pa}[1]{\left(#1\right)}
\newcommand{\paq}[1]{\left[#1\right]}
\newcommand{\pag}[1]{\left\{#1\right\}}
\newcommand{\flux}{F}
\newcommand{\dpa}{\partial}
\newcommand{\K}{{\bf k}}
\newcommand{\Q}{{\bf q}}
\newcommand{\X}{{\bf x}}
\newcommand{\ccdot}{.}
\newcommand{\vep}{\varepsilon}
\newcommand{\z}[1]{\zeta_{#1}}
\newcommand{\ppadded}[1]{{\color{blue}[Pp : #1]}}
\newcommand{\added}[1]{{\color{red}[added : #1]}}
\newcommand{\rs}[1]{{\color{brown}[RS : #1]}}
\newcommand{\ste}[1]{{\color{red}[SF: #1]}}
\newcommand{\cs}[1]{{\color{blue}{CS: #1}}}
\allowdisplaybreaks


\title{\boldmath Static Two-Body Potential at Fifth Post-Newtonian Order}

\author{
Stefano Foffa$^a$, 
Pierpaolo Mastrolia$^b$, 
Riccardo Sturani$^c$, 
Christian Sturm$^d$,
William J. Torres Bobadilla$^e$
}
\affiliation{
$^a$D\'epartement de Physique Th\'eorique and Centre for Astroparticle Physics, Universit\'e de 
             Gen\`eve, CH-1211 Geneva, Switzerland \\
$^b$Dipartimento di Fisica ed Astronomia, Universit\`a di
  Padova, Via Marzolo 8, 35131 Padova, Italy\\
  INFN, Sezione di Padova, Via Marzolo 8, 35131 Padova, Italy \\
$^c$ International Institute of Physics (IIP),
Universidade Federal do Rio Grande do Norte (UFRN) CP 1613, 59078-970
Natal-RN Brazil \\ 
$^d$Universit{\"a}t W{\"u}rzburg, Institut f{\"u}r Theoretische Physik und Astrophysik, Emil-Hilb-Weg 22, D-97074 W{\"u}rzburg, Germany\\
$^e$Instituto de F\'{\i}sica Corpuscular, Universitat de Val\`{e}ncia -- Consejo Superior de Investigaciones Cient\'{\i}ficas, 
Parc Cient\'{\i}fic, E-46980 Paterna, Valencia, Spain.}

\begin{abstract}
We determine the gravitational interaction between two compact bodies
up to the sixth power in Newton's constant $G_N$, in the static limit.
This result is achieved within the effective field theory approach
to General Relativity, 
and exploits a manifest factorization property of static diagrams which allows
us to
derive static post Newtonian (PN) contributions of $(2n+1)$-order 
in terms of lower order ones.
We recompute in this fashion the 1PN and 3PN static potential, and
present the novel 5PN contribution. 
\end{abstract}

\maketitle

\flushbottom



\section{Introduction}
\label{sec:intro}
The study of the conservative dynamics of the two-body problem in
General Relativity (GR)
is one of the pillars which allows us to determine the gravitational waveform templates 
for the LIGO/Virgo \cite{TheLIGOScientific:2014jea,TheVirgo:2014hva}
data analysis pipeline \cite{Taracchini:2012ig,Schmidt:2014iyl}. 
The future generation of detectors such as the Einstein Telescope
\cite{Punturo:2010zz} and LISA \cite{Audley:2017drz} 
are expected to gain in sensitivity at least one order of magnitude
with respect to the current generation
of ground-based interferometers. Therefore, more accurate predictions on the
theoretical side will be required for Gravitational Wave (GW) astrophysics in the
next decade \cite{Lindblom:2008cm,Antonelli:2019ytb}.

Deviations from the Newton potential due to GR effects can be studied in
the so-called post-Newtonian (PN) framework, that is by expanding in powers of the two
virial-related quantities, such as the compactness $R_S/r\sim G_N m/r$ and the
relative (squared) velocity $v^2\sim G_N m/r$, 
where $R_S$, $m$, $r$ and $G_N$ are \ the Schwarzschild radius of the
system, its mass and size, and Newton's constant, respectively. 
The first complete 1PN computation was preformed by Einstein, Infeld and Hoffmann in \cite{Einstein:1938yz};
since then, the evaluation of the higher-order terms has been a formidable
effort, whose current state of the art,
after the calculation of the 2PN \cite{Damour:1982wm,Damour:1985mt} and 3PN \cite{Damour:2001bu,Blanchet:2003gy,Itoh:2003fy} contributions,
is represented by the determination of
the energy at 4PN order, which was achieved for the first time in \cite{Damour:2014jta,Damour:2015isa,Damour:2016abl}
and later confirmed in \cite{Bernard:2015njp,Bernard:2016wrg,Bernard:2017bvn,Marchand:2017pir,Bernard:2017ktp}
and in \cite{Foffa:2012rn,Foffa:2016rgu,Foffa:2019rdf,Foffa:2019yfl}.

The next complexity level, namely the fifth post-Newtonian approximation
(5PN), is
qualitatively important because of the first appearance of spin-independent
finite size corrections.
Several partial results towards this challenging precision level
have become recently available 
in the so-called post-Minkowskian expansion, {\it i.e.} the  
 expansion in $G_N$ only, for any given order in $v$, 
up to the third order in Newton's constant $G_N$
\cite{Ledvinka:2008tk,Foffa:2013gja,Blanchet:2018yvb,Damour:2017zjx,Cheung:2018wkq,Bern:2019nnu}.

In the present Letter, we provide a novel contribution to the 5PN
dynamics by tackling the determination of the highest possible power
in $G_N$, namely $G_N^6$ at 5PN, which amounts 
to determine the potential in the {\it static limit}. \\
This goal is achieved by building on the ideas and the method of \cite{Foffa:2016rgu}, where we computed the static potential at
4PN by adopting the effective field theory (EFT) approach to GR
\cite{Goldberger:2004jt,Goldberger:2007hy,Foffa:2013qca,Rothstein:2014sra,Porto:2016pyg},
in combination with techniques for the evaluation of multiloop scattering amplitudes in momentum space
(see also \cite{Damour:2017ced} for a related computation in direct space).
The computation of 5PN static corrections turns out to be actually less demanding than
the corresponding 4PN ones, owing to a 
factorization property of the static contributions, yielding a drastic
simplification at {\it odd}-PN orders, which is explicit and intuitive in the EFT
approach, and is formalized in the current Letter.

The $G_N^6$ subsector computed here is the highest order ever
  computed in powers of $G_N$ and, if done by brute force, involves the evaluation of the most
  complex integrals present at 5PN, which, within the EFT expansion,
 come from the graphs with the largest loop number. Therefore, its
  determination,
  presented here for the first time, paves the way to the completion
  of 5PN corrections, and confirms that the adopted methods
  are suitable to systematically tackle higher PN-order computations.

\section{Effective Field Theory approach}

The evaluation of post-Newtonian corrections to the dynamics of binary
systems can be addressed within the by now established EFT framework
\cite{Goldberger:2004jt}, reviewed in \cite{Foffa:2013qca,Porto:2016pyg,Levi:2018nxp}.
Following the lines
and notations of \cite{Foffa:2011ub, Foffa:2012rn}, we consider
the action of the system, given by
\be
\label{action}
S = S_{\rm pp} + S_{\rm bulk}
\ee
in terms of the world-line point particle action, representing the binary components
(for spinless point masses and neglecting tidal effects)
 \be
\label{az_pp}
S_{\rm pp}=-\!\!\sum_{i=1,2} \!\!m_i\!\!\int {\rm d}\tau_i 
 = -\!\!\sum_{i=1,2} \!\!m_i\!\!\int\!\!\sqrt{-g_{\mu\nu}(x_i) {\rm
     d}x_i^\mu {\rm d}x_i^\nu}\,,
\quad
\ee
and of the canonical Einstein-Hilbert action 
plus a gauge-fixing (harmonic condition) term ~\cite{Blanchet:2014,Bernard:2015njp},
\begin{eqnarray}
\label{az_bulk}
S_{\rm bulk} 
&=& S_{\rm EH}+ S_{\rm GF} \nonumber \\ 
&=& 2 \Lambda^2 \int {\rm d}^{d+1}x\sqrt{-g}\left[ R(g)-\frac{1}{2}\Gamma_\mu\Gamma^\mu\right]\,,
\end{eqnarray}
where $\Gamma^\mu \equiv g^{\rho\sigma}\Gamma^\mu_{\rho\sigma}$.
In the above formula, $\Lambda^{-2}\equiv 32 \pi G_N L^{d-3}$, where 
$G_N$ is the three-dimensional Newton constant, 
and $L$ is an arbitrary length scale that keeps the correct dimensions of $\Lambda$ in dimensional regularization, which cancels out in the expression of physical observables. In this framework, a Kaluza-Klein (KK) 
parametrization  of the metric \cite{Blanchet:1989ki,Kol:2007bc,Kol:2007rx} is usually
adopted:
\be
\label{met_nr}
g_{\mu\nu}=e^{2\phi/\Lambda}\pa{
\ba{cc}
-1 & A_j/\Lambda \\
A_i/\Lambda &\quad e^{- c_d\phi/\Lambda}\gamma_{ij}-
A_iA_j/\Lambda^2\\
\ea
}\,,
\ee
with $\gamma_{ij} \equiv \delta_{ij}+\sigma_{ij}/\Lambda$,
$c_d=2(d-1)/(d-2)$ and $i,j$ running over the $d$ spatial dimensions.
Accordingly, the degrees of freedom of the graviton field are
reparametrized in terms of a scalar field $\phi$, a vector field
$ A_i$, and a symmetric tensor field $\sigma_{ij}$.
The field $A_i$ is not actually needed in the static limit because it
always comes in association with the velocity of one of the compact
bodies, so it will henceforth be set to zero.

In terms of the metric parametrization eq.~(\ref{met_nr}), with $A_i=0$,
each world-line coupling to the gravitational degrees of freedom
$\phi$, $\sigma_{ij}$  reads
\renewcommand{\arraystretch}{1.4}
\be
\label{matter_grav}
S_{pp}&=&\ds-m\int {\rm d}t\ e^{\phi/\Lambda}\sqrt{1
-e^{-c_d \phi/\Lambda}\pa{v^2+\frac{\sigma_{ij}}{\Lambda}v^iv^j}}
\nonumber \\ 
&\stackrel[\mbox{\tiny{static}}]{}{\longrightarrow}& -m \int {\rm d}t\ e^{\phi/\Lambda}\,,
\ee
\renewcommand{\arraystretch}{1.4}
and its Taylor expansion provides the various particle-gravity
vertices involving $\phi$, like 
the coupling of $\phi$ to matter fields,  i.e. the $m\phi^n$-vertex,

\begin{eqnarray}
\parbox{20mm}{
\begin{tikzpicture}[line width=1 pt,node distance=0.4 cm and 0.4 cm]
\coordinate[] (v1);
\coordinate[right = of v1] (v2);
\coordinate[right = of v2] (v3);
\coordinate[right = of v3] (v4);
\coordinate[right = of v4] (v5);
\coordinate[below = of v3] (v9);
\coordinate[below = of v9, label=center: $\overset{\cdots}{n}$] (v6);
\coordinate[left = of v6] (v7);
\coordinate[right = of v6] (v8);
\draw[phi] (v3) -- (v7);
\draw[phi] (v3) -- (v8);
\draw[fermionnoarrow] (v1) -- (v5);
\end{tikzpicture}
}
\!\!\!
&=& \ 
-\frac{\text{i}\ m}{n!  \Lambda^n} \ , 
\end{eqnarray}
where black lines stands for matter and dashed blue lines indicate
$\phi$ modes. 
Also the pure gravity sector $S_{\rm bulk}$ can be explicitly written
in terms of the KK variables; for the purpose of this Letter, it is sufficient to report here only the structure of the static terms
not containing the field $\vec{A}$~\footnote{It is understood that spatial indices in this expression are contracted by means of the spatial metric $\gamma_{ij}$, which implies the
appearance of extra $\sigma$ fields.
}:
\begin{eqnarray}
\label{bulk_action}
\ds S_{\rm bulk} &\supset& \ds \int {\rm d}^{d+1}x\sqrt{\gamma}
\left\{ f(\sigma_{ij})- c_d (\vec{\nabla}\phi)^2\right\}\,,
\end{eqnarray}
\renewcommand{\arraystretch}{1.}
where $f$ is a function depending on the field $\sigma_{ij}$ only.
The complete set of Feynman rules, also involving the fields $\sigma_{ij}$
and $A_i$
(respectively indicated by green and red lines), can be found in \cite{Foffa:2016rgu}.

The two-body effective action can be found by integrating out the gravity fields from the above-derived actions
\begin{equation}
\label{eq:seff}
 \exp[\text{i} S_{\rm eff}]=\int D\phi D\sigma_{ij} \exp[\text{i}(S_{\rm bulk}+ S_{\rm pp})]\,.
\end{equation}
Within the field-theoretical approach, the functional integration can be perturbatively 
expanded in terms of Feynman diagrams involving the gravitational 
degrees of freedom as internal lines only,
viewed as dynamical fields emitted and absorbed by the point particles which 
are taken as nondynamical sources.
Each diagram shows a manifest power counting both in the bodies' relative velocities and in $G_N$
(any bulk vertex involving $k$ fields carries a factor
$(G_N)^{{k\over 2}-1}$
and any $m\phi^n$-vertex
carries a factor 
$(G_N)^{{n \over 2}}$ ),
thus allowing for a systematic PN classification.
%
The most elementary diagram in the EFT approach is represented by 
the Newton-potential graph

\begin{eqnarray}
{\cal V}_N = \
\parbox{15mm}{
\begin{tikzpicture}[line width=1 pt,node distance=0.5 cm and 0.3125 cm]
\coordinate[] (v1);
\coordinate[right = of v1] (v2);
\coordinate[right = of v2] (v3);
\coordinate[below = of v2] (v4);
\coordinate[below = of v4] (v5);
\coordinate[left = of v5] (v6);
\coordinate[right = of v5] (v7);
\draw[phi] (v2) -- (v5);
\draw[fermionnoarrow] (v1) -- (v3);
\draw[fermionnoarrow] (v6) -- (v7);
\end{tikzpicture}
}\!\!\!\!\!\!\!\!\!\!\!\!
= -\frac{G_N m_1 m_2}{r} \ ,
\end{eqnarray}
na\"\i vely dubbed as 0PN diagram.

\section{Factorization theorem}
\begin{figure}[h]
\begin{center}
\includegraphics[width=.3\linewidth]{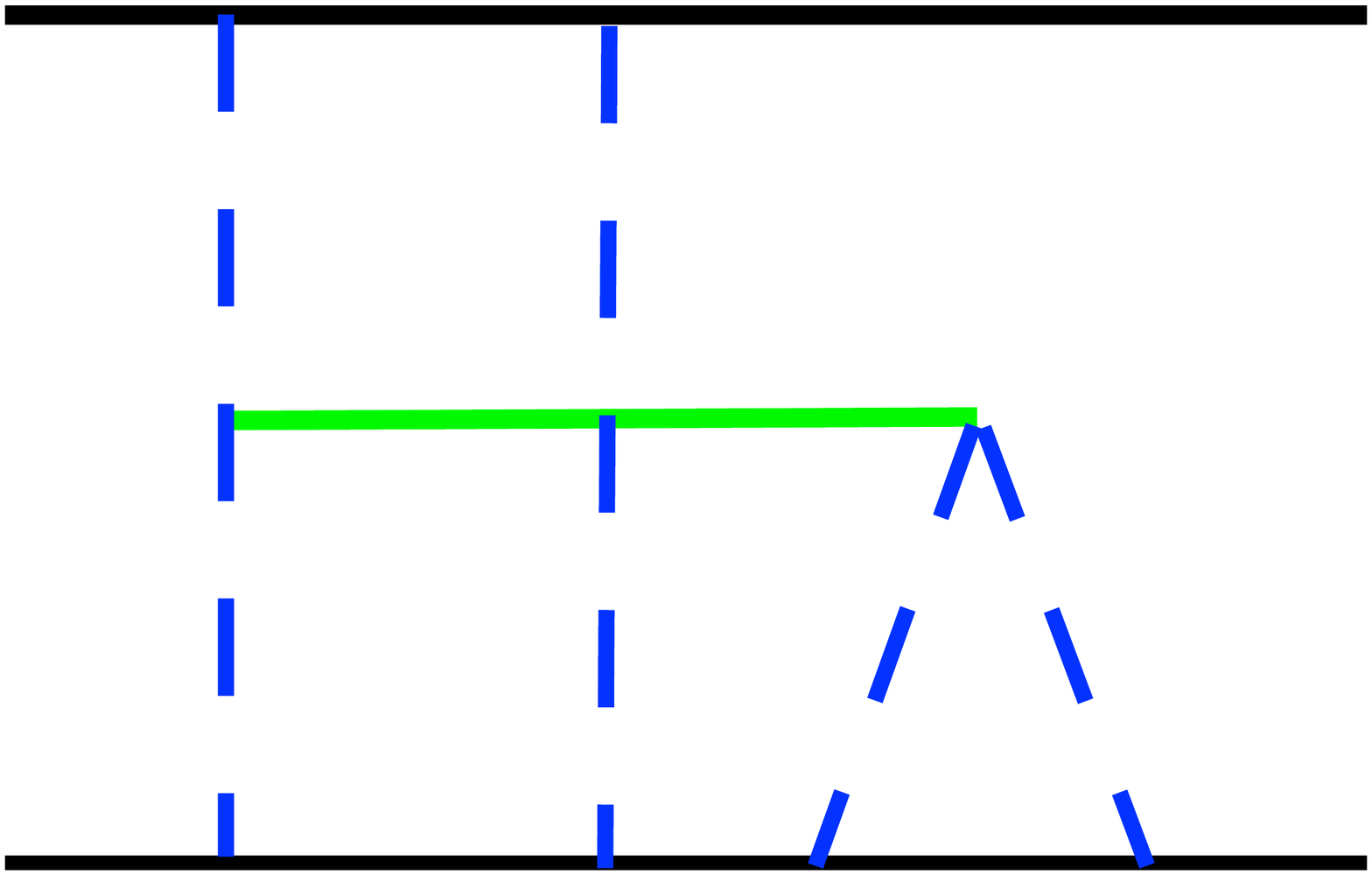}\hspace{1cm}
\includegraphics[width=.3\linewidth]{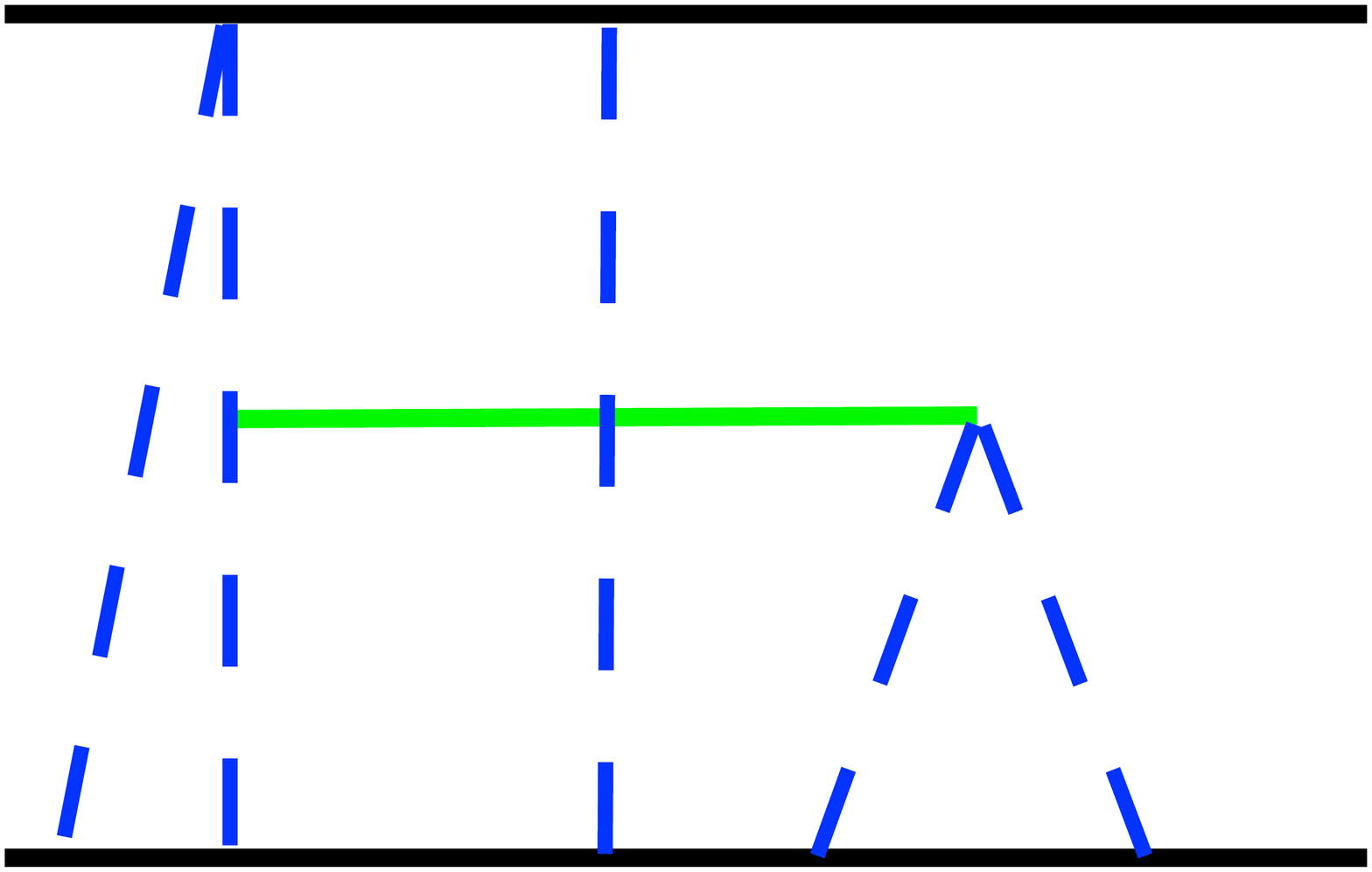}
\caption{Examples of a prime 4PN-graph (left) and of a factorizable
  5PN-graph (right): the latter can be obtained by sewing the former
  and the Newton potential diagram. 
}
\label{fig:prime}
\end{center}
\end{figure}

{\bf Definition:} 
Static EFT-gravity diagrams can be classified according to the type of couplings
between matter and $\phi$ fields. We can distinguish between:
 {\it factorizable graphs}, which
contain at least one $m\phi^n$-vertex with $n>1$, and 
{\it prime graphs}, which contain only matter-$\phi$ vertices of the type $m\phi$, 
namely where each $\phi$, coming from the bulk (and not propagating
between bulk vertices), couples individually 
to matter (see fig. \ref{fig:prime} left).

Factorizable graphs can be obtained by {\it sewing} together 
two, or more, subgraphs that, upon merging, share
a $m\phi^n$-vertex ($n>1$)
(see fig. \ref{fig:prime} right).

{\bf Proposition:}
Inspection of eq.(\ref{bulk_action}) shows that the only bulk gravity
vertices allowed in a static graph are those containing {\it (a)}
zero or two $\phi$s and 
{\it (b)} any number of $\sigma_{ij}$s;
the latter cannot however be attached to any particle, see
eq.(\ref{matter_grav}), so they can just propagate between bulk
vertices. This observation is crucial to prove an important property of prime
graphs, which constitute the first novel result of this communication:\\

{\bf Theorem:}  Static prime graphs exist only at even $2n$-PN 
orders. Equivalently, static graphs at odd $(2n+1)$-PN orders are factorizable.

{\it Proof:} 
This statement can be proven by showing that any prime static 
graph must have an even number of $\phi$ fields attached to the particles.

For the Newtonian graph, it is trivially true by construction. 
Graphs generated by PN corrections, ${\cal O}(G_N^2)$, necessarily
contain bulk vertices
$\phi\phi\sigma^k$ (with $k\geq1$), coming from the expansions of the graviton self-interaction terms. For these diagrams, two cases may occur:
{\it i)} each internal $\phi$ propagator is contracted on the one side
with a matter-$\phi$ vertex, 
and, on the other side, with a $\phi\phi\sigma^k$ vertex,
therefore it contributes with one power of $m_i$ to the
mass dimensions of the graph;
{\it ii)} a $\phi$ propagator, not coupled with matter, 
must necessarily connect two $\phi\phi \sigma^k$ vertices, 
therefore it does not contribute to the mass dimensions of the graph.
Since the bulk vertices between $\phi$ and $\sigma$ fields
($\phi\phi\sigma$, $\phi\phi\sigma\sigma$, \dots) are quadratic in $\phi$, 
and because prime graphs are characterized by either {\it (i)} or {\it (ii)} , 
we can conclude that  
the {\it total} number
of $\phi$ fields that depart from the bulk vertices and couple to
matter (either $m_1$ or $m_2$)
is an {\it even} number. 

This implies that, being $n_i$ the number of $\phi$ fields coupled to
the matter $m_i$ $(i=1,2)$, the total masslike power of static prime graphs is
$m_1^{n_1} m_2^{n_2}$, with $n_1+n_2 = 2n$ and $n \in {\mathbb
  N}^{+}$. On the other side, they correspond to static classical
contributions, 
therefore, they must consequently scale as $G_N^{(2n-1)} m_1^{n_1} m_2^{n_2}/r^{(2n-1)}$
(classical diagrams do not contain loops in the dynamical fields),
finally implying that they belong to an even-PN order.
\hfill $\square$\\

Due to the factorization theorem, the general structure of the 
contribution to the potential of a given $n$-PN {\it factorizable} diagram, in terms
of the product of lower PN-order graphs, reads
\begin{eqnarray}
{\cal V}_{n}^{\rm factorizable} = \Big({\cal V}_{L,n_1} \times {\cal
  V}_{R,n_2}\Big) \times {\cal K}\times{\cal C}\,,
\end{eqnarray}
where:
{\it i)} the PN orders, $n_1$ of the left graph ${\cal V}_{L}$ and $n_2$
of the right graph ${\cal V}_{R}$, are such that
$n_1+n_2+1 = n$;
{\it ii)} ${\cal K}$ accounts for the new
matter-$\phi^k$ vertex of ${\cal V}_{n}$ (emerging from the sewing) out of the ones included in the lower order contributions,
${\cal V}_{L,n_1}$ and $ {\cal V}_{R,n_2}$; and {\it iii)} ${\cal C}=C^{\rm factorizable}_n/(C_{L,n_1} \times C_{R,n_2})$ where the $C$'s are the combinatoric
factors associated with each graph.

\section{Gravity and Field Theory Diagrams}

In a quantum field theory approach, any EFT-gravity graph can
be interpreted as four-particle scattering amplitude \cite{Foffa:2016rgu}.
The contribution of each amplitude to the two-body potential ${\cal V}$ can be
obtained by taking its Fourier transform, 
\begin{equation}
{\cal V}
\quad =  \text{i}  \lim_{d \to 3} \ 
\int_p \ e^{\text{i} p \cdot r}
\parbox{25mm}{
\begin{tikzpicture}[line width=1 pt,node distance=0.4 cm and 0.4 cm]
\coordinate[label=left: \tiny$2$] (v1);
\coordinate[right = of v1] (v2);
\coordinate[right = of v2] (v3);
\coordinate[right = of v3] (v4);
\coordinate[right = of v4, label=right: \tiny$3$] (v5);
\coordinate[below = of v1] (v6);
\coordinate[below = of v6, label=left: \tiny$1$] (v7);
\coordinate[right = of v7] (v8);
\coordinate[right = of v8] (v9);
\coordinate[right = of v9] (v10);
\coordinate[right = of v10, label=right: \tiny$4$] (v11);
\fill[black!25!white] (v8) rectangle (v4);
\draw[fermionnoarrow] (v1) -- (v5);
\draw[fermionnoarrow] (v7) -- (v11);
\end{tikzpicture}
}
\end{equation}
where, $\int_p \equiv \int {{\rm d}^dp /(2 \pi)^d}$,
the box diagram stands for a generic EFT-gravity diagram,
and $p$ is the momentum transfer of the source 
(assuming momentum conservation $p_1+p_2 = p_3+p_4$, then  
$p=p_3-p_2=p_1-p_4$).
Since the sources, represented by black lines, are static and do not propagate, 
any EFT-gravity amplitude at order $G_N^\ell$ can be
mapped into an $(\ell-1)$ loop two-point function with massless internal
lines and external momentum $p$ $(p^2 \ne 0)$ 
\cite{Foffa:2016rgu}. This observation was crucial to perform
the 4PN static calculation by employing computational techniques
developed for the evaluation of multiloop Feynman integrals in high-energy particle
physics.  Moreover, in the current Letter, we observe that the integration on $p$ 
can be seen as an additional loop integration, hence it can be represented by an
$\ell$-loop vacuum diagram, obtained by joining the external legs into a
propagatorlike line (indicated by an inner black line), as  
\begin{eqnarray}
\int_p \ e^{\text{i} p \cdot r} \
\parbox{15mm}{
\begin{tikzpicture}[line width=1 pt,node distance=0.4 cm and 0.4 cm]
\coordinate[] (v1);
\coordinate[below= of v1] (v2);
\coordinate[below= of v2] (v3);
\coordinate[below= of v3] (v4);
\coordinate[below= of v4] (v5);
\draw[fermionnoarrow] (v1) -- (v2);
\draw[fermionnoarrow] (v4) -- (v5);
\fill[black!25!white] (v3) circle (0.4cm);
\end{tikzpicture}
}
\!\!\!\!\!\!
\equiv\
\parbox{15mm}{
\begin{tikzpicture}[line width=1 pt,node distance=0.4 cm and 0.4 cm]
\coordinate[] (v2);
\coordinate[below= of v2] (v3);
\coordinate[below= of v3] (v4);
\coordinate[right= of v3] (v4);
\fill[white!25!white,draw=black] (0.24,-0.4) ellipse (0.32cm and 0.44cm);
\fill[black!25!white] (v3) circle (0.4cm);
\end{tikzpicture}
}
\!\!\!
\to \
\parbox{15mm}{
\begin{tikzpicture}[line width=1 pt,node distance=0.4 cm and 0.4 cm]
\coordinate[] (v2);
\coordinate[below= of v2] (v3);
\coordinate[below= of v3] (v4);
\coordinate[right= of v3] (v4);
\fill[black!25!white] (v3) circle (0.4cm);
\fill[white!25!white] (0.2,-0.4) circle (0.2cm);
\draw[fill=black] (v4) circle (.05cm);
\end{tikzpicture}
}
\!\!\!\!. 
\end{eqnarray}
In the last step, we introduce a suggestive diagrammatic representation of the
Fourier integral as an $\ell$-loop vacuum graph by {\it pinching} 
the internal black line. The presence of the {\it dot} ``$\bullet$''
 indicates the residual $r$ dependence of the contribution
(not to be confused by fully massless, hence scaleless vacuum
diagrams that vanish in
dimensional regularization).

In the case of factorizable
EFT-diagrams, the pinching generates the product of factorized vacuum
diagrams. For example, the contribution to the 5PN potential of the
diagram in fig.~\ref {fig:prime} (right) becomes
\begin{equation}
\int_p \ e^{\text{i} p \cdot r} \
\parbox{15mm}{
\begin{tikzpicture}[line width=1 pt,node distance= 0.6 cm and 0.3 cm]
\coordinate[] (v0);
\coordinate[right = of v0] (v1);
\coordinate[right = of v1] (v2);
\coordinate[right = of v2] (v2a);
\coordinate[right = of v2a] (v6);
\coordinate[right = of v6] (v7);
\coordinate[above right = of v7] (v9);
\coordinate[below right = of v9] (v8);
\coordinate[right = of v8] (v12);
\coordinate[above = of v2] (v3);
\coordinate[above = of v3] (v4);
\coordinate[left = of v4] (v4a);
\coordinate[left = of v4a] (v11);
\coordinate[right = of v4] (v4b);
\coordinate[right = of v4b] (v5);
\coordinate[above right = of v9] (v9a);
\coordinate[right = of v9a] (v10);
\draw[phi] (v7) -- (v9);
\draw[phi] (v8) -- (v9);
\draw[phi] (v5) -- (v6);
\draw[phi] (v4) -- (v2);
\draw[phi] (v4) -- (v1);
\draw[sigma] (v3) -- (v9);
\draw[fermionnoarrow] (v0) -- (v12);
\draw[fermionnoarrow] (v11) -- (v10);
\end{tikzpicture}
}
\qquad\quad\to\quad
\parbox{15mm}{
\begin{tikzpicture}[line width=1 pt,node distance= 0.3 cm and 0.3 cm]
\coordinate[] (v0);
\coordinate[above = of v0] (v1);
\coordinate[below = of v0] (v2);
\coordinate[left = of v0] (v3a);
\coordinate[left = of v3a] (v3);
\coordinate[right = of v0] (v4a);
\coordinate[right = of v4a] (v4);
\draw[blue,dashed] (v0) circle (.6cm);
\draw[sigma] (v1) -- (v3);
\draw[sigma] (v2) -- (v3);
\draw[phi] (v4) arc (30:100:0.6cm);
\draw[phi] (v4) arc (-30:-100:0.6cm);
\draw[phi] (v1) arc (210:280:0.6cm);
\draw[phi] (v2) arc (-210:-280:0.6cm);
\draw[fill=black] (v4) circle (.05cm);
\draw[phi] (v4) arc (180:-180:0.225cm);
\end{tikzpicture}
}
\quad\ ,
\end{equation}
directly representing the product of the Newton potential and of a
4PN-term,
respectively represented by a one-loop and a five-loop vacuum diagram.

\section{Results}

We verified that the static potential at 1PN and 3PN \cite{Foffa:2011ub} can be derived by
applying the factorization theorem to the relevant diagrams.




 We now apply the factorization theorem to the 5PN case, computed for
 the first time in the present Letter.
There are 154 diagrams to evaluate and it is convenient to divide them
in four classes, according to their factorization patterns.
For ease of notation, we arrange the 5PN static graphs in four
subsets, displaying the lower-PN corrections they stem out from, 
and give the factors ${\cal K}$ and ${\cal C}$ as understood.

\noindent
1. 
There are 11 diagrams  composed of six Newtonian factors, combined in
different ways, and schematically represented as
\begin{eqnarray}
\left(\,\parbox{15mm}{
\begin{tikzpicture}[line width=1 pt,node distance=0.5 cm and 0.3125 cm]
\coordinate[] (v1);
\coordinate[right = of v1] (v2);
\coordinate[right = of v2] (v3);
\coordinate[below = of v2] (v4);
\coordinate[below = of v4] (v5);
\coordinate[left = of v5] (v6);
\coordinate[right = of v5] (v7);
\draw[fermionnoarrow] (v1) -- (v3);
\draw[phi] (v2) -- (v5);
\draw[fermionnoarrow] (v6) -- (v7);
\end{tikzpicture}
}\!\!\!\!\!\!\!\!\!\!\!\!\right)^6 \ .
\end{eqnarray}
The contribution to the 5PN potential coming from this set of diagrams is
\begin{eqnarray}
 {\cal V}_{N^6} &=&
\frac{1}{720}\frac{G_N^6 m_1^6 m_2}{r^6}+\frac{1}{3}\frac{G_N^6 m_1^5
                    m_2^2}{r^6} \nonumber \\
& &
+3\frac{G_N^6 m_1^4 m_2^3}{r^6}+(m_1\leftrightarrow m_2)\,.
\end{eqnarray}

\noindent
2.
One can build static factorizable diagrams as products of three
Newtonian graphs, and either of the 2PN prime graphs, schematically
represented as:
\begin{equation}
\left(\ \parbox{15mm}{
\begin{tikzpicture}[line width=1 pt,node distance=0.5 cm and 0.3125 cm]
\coordinate[] (v1);
\coordinate[right = of v1] (v2);
\coordinate[right = of v2] (v3);
\coordinate[below = of v2] (v4);
\coordinate[below = of v4] (v5);
\coordinate[left = of v5] (v6);
\coordinate[right = of v5] (v7);
\draw[fermionnoarrow] (v1) -- (v3);
\draw[phi] (v2) -- (v5);
\draw[fermionnoarrow] (v6) -- (v7);
\end{tikzpicture}
}\!\!\!\!\!\!\!\!\!\!\!\!\right)^3
\times
\left(\
\parbox{15mm}{
\begin{tikzpicture}[line width=1 pt,node distance=0.5 cm and 0.3125 cm]
\coordinate[] (v1);
\coordinate[right = of v1] (v2);
\coordinate[right = of v2] (v3);
\coordinate[right = of v3] (v4);
\coordinate[right = of v4] (v5);
\coordinate[below = of v1] (v6);
\coordinate[below = of v6] (v7);
\coordinate[right = of v7] (v8);
\coordinate[right = of v8] (v9);
\coordinate[right = of v9] (v10);
\coordinate[right = of v10] (v11);
\coordinate[below = of v2] (v12);
\coordinate[below = of v4] (v13);
\draw[sigma] (v12) -- (v13);
\draw[phi] (v2) -- (v8);
\draw[phi] (v4) -- (v10);
\draw[fermionnoarrow] (v1) -- (v5);
\draw[fermionnoarrow] (v7) -- (v11);
\end{tikzpicture}
}
\parbox{15mm}{
\begin{tikzpicture}[line width=1 pt,node distance=0.333 cm and 0.3125 cm]
\coordinate[] (v1);
\coordinate[right = of v1] (v2);
\coordinate[right = of v2] (v3);
\coordinate[right = of v4] (v5);
\coordinate[below right = of v2] (v6);
\coordinate[below = of v6] (v7);
\coordinate[below left = of v7] (v8);
\coordinate[below right = of v7] (v9);
\coordinate[left = of v8] (v10);
\coordinate[right = of v9] (v11);
\draw[sigma] (v6) -- (v7);
\draw[phi] (v2) -- (v6);
\draw[phi] (v4) -- (v6);
\draw[phi] (v8) -- (v7);
\draw[phi] (v9) -- (v7);
\draw[fermionnoarrow] (v1) -- (v5);
\draw[fermionnoarrow] (v10) -- (v11);
\end{tikzpicture}
}
\parbox{15mm}{
\begin{tikzpicture}[line width=1 pt,node distance=0.5 cm and 0.25 cm]
\coordinate[] (v1);
\coordinate[right = of v1] (v2);
\coordinate[below = of v2] (v3);
\coordinate[below = of v3] (v4);
\coordinate[right = of v3] (v5);
\coordinate[right = of v5] (v6);
\coordinate[below left = of v6] (v7);
\coordinate[below right = of v6] (v8);
\coordinate[right = of v8] (v9);
\coordinate[left = of v4] (v10);
\coordinate[above right= of v6] (v11);
\coordinate[right= of v11] (v12);
\draw[phi] (v2) -- (v4);
\draw[phi] (v7) -- (v6);
\draw[phi] (v8) -- (v6);
\draw[sigma] (v3) -- (v6);
\draw[fermionnoarrow] (v9) -- (v10);
\draw[fermionnoarrow] (v1) -- (v12);
\end{tikzpicture}
} \right)\ .
\end{equation}
This set contains 49 diagrams, 9 of which are vanishing, 
because one of the 2PN factors is indeed zero. 
The combined contribution of the remaining diagrams is
\begin{eqnarray}
 {\cal V}_{N^3 \times {\rm 2PN}} &=&
\frac{1}{18}\frac{G_N^6 m_1^6 m_2}{r^6}+\frac{16}{3}\frac{G_N^6 m_1^5
                             m_2^2}{r^6}
\nonumber \\
& &
+\frac{229}{6}\frac{G_N^6 m_1^4 m_2^3}{r^6}+(m_1\leftrightarrow m_2)\,.
\end{eqnarray}

\noindent 
3. In this class, we consider 5PN diagrams schematically represented by
the product of one Newtonian graph with each of the
  $25$ static prime 4PN diagrams studied in \cite{Foffa:2016rgu}
  (the cardinal number attached to each graph is the
  same as in \cite{Foffa:2016rgu}, for ease of comparison)


\begin{equation}
\parbox{15mm}{
\begin{tikzpicture}[line width=1 pt,node distance=0.5 cm and 0.3125 cm]
\coordinate[] (v1);
\coordinate[right = of v1] (v2);
\coordinate[right = of v2] (v3);
\coordinate[below = of v2] (v4);
\coordinate[below = of v4] (v5);
\coordinate[left = of v5] (v6);
\coordinate[right = of v5] (v7);
\draw[fermionnoarrow] (v1) -- (v3);
\draw[phi] (v2) -- (v5);
\draw[fermionnoarrow] (v6) -- (v7);
\end{tikzpicture}
}\!\!\!\!\!\!\!\!\!\!\!
\times\
\left(\
\parbox{15mm}{
\begin{tikzpicture}[line width=1 pt,node distance= 0.5 cm and 0.25 cm]
\coordinate[] (v1);
\coordinate[right = of v1] (v2);
\coordinate[above right = of v2,label= above left :\tiny$26$] (v3);
\coordinate[below right = of v3] (v4);
\coordinate[right = of v4] (v6);
\coordinate[above = of v6] (v7);
\coordinate[above = of v7] (v8);
\coordinate[right = of v6] (v9);
\coordinate[above right = of v9] (v10);
\coordinate[below right = of v10] (v11);
\coordinate[right = of v11] (v12);
\coordinate[above = of v1] (v13);
\coordinate[above = of v13] (v14);
\coordinate[above = of v12] (v15);
\coordinate[above = of v15] (v16);
\draw[phi] (v2) -- (v3);
\draw[phi] (v3) -- (v4);
\draw[phi] (v6) -- (v8);
\draw[phi] (v9) -- (v10);
\draw[phi] (v10) -- (v11);
\draw[sigma] (v3) -- (v10);
\draw[fermionnoarrow] (v1) -- (v12);
\draw[fermionnoarrow] (v14) -- (v16);
\end{tikzpicture}
}\qquad\
\boldsymbol{\cdots}\quad
\parbox{15mm}{
\begin{tikzpicture}[line width=1 pt,node distance= 0.25 cm and 0.25 cm]
\coordinate[] (v1);
\coordinate[right = of v1] (v2);
\coordinate[right = of v2] (v3);
\coordinate[right = of v3] (v4);
\coordinate[below = of v4, label=  right :\tiny$50$] (v5);
\coordinate[below left= of v5] (v6);
\coordinate[left= of v6] (v7);
\coordinate[left= of v7] (v8);
\coordinate[below right= of v7] (v6b);
\coordinate[below= of v6b] (v9);
\coordinate[right= of v9] (v10);
\coordinate[left= of v9] (v11);
\coordinate[left= of v11] (v12);
\coordinate[left= of v1] (v0);
\coordinate[left= of v12] (v13);
\coordinate[right= of v10] (v14);
\coordinate[right= of v4] (v15);
\draw[sigma] (v8) -- (v7);
\draw[sigma] (v5) -- (v7);
\draw[sigma] (v6b) -- (v7);
\draw[phi] (v3) -- (v9);
\draw[phi] (v10) -- (v4);
\draw[phi] (v12) -- (v1);
\draw[fermionnoarrow] (v14) -- (v13);
\draw[fermionnoarrow] (v0) -- (v15);
\end{tikzpicture}
}
\!\right)\ .
\end{equation}


This set contains 79 diagrams, 16 of which are vanishing (due to
vanishing 4PN factors). The remaining 63 diagrams give
\begin{eqnarray}
 {\cal V}_{N \times {\rm 4PN}} &=&
\frac{1}{5}\frac{G_N^6 m_1^6 m_2}{r^6}+\frac{23}{3}\frac{G_N^6 m_1^5
                            m_2^2}{r^6}
\nonumber \\
& &
+\frac{166}{3}\frac{G_N^6 m_1^4 m_2^3}{r^6}+(m_1\leftrightarrow m_2)\,.
\end{eqnarray}
Interestingly, let us observe that 
although this set contains contributions that
are individually divergent in the $d \to 3$ limit, as well as factors of $\pi^2$, within their sum all poles
and irrational factors cancel, and the result is indeed finite and rational.

\noindent
4.
Finally, we consider static 5PN diagram formed by the product of two 2PN graphs,
schematically represented as
\begin{equation}
\left(\
\parbox{15mm}{
\begin{tikzpicture}[line width=1 pt,node distance=0.5 cm and 0.3125 cm]
\coordinate[] (v1);
\coordinate[right = of v1] (v2);
\coordinate[right = of v2] (v3);
\coordinate[right = of v3] (v4);
\coordinate[right = of v4] (v5);
\coordinate[below = of v1] (v6);
\coordinate[below = of v6] (v7);
\coordinate[right = of v7] (v8);
\coordinate[right = of v8] (v9);
\coordinate[right = of v9] (v10);
\coordinate[right = of v10] (v11);
\coordinate[below = of v2] (v12);
\coordinate[below = of v4] (v13);
\draw[sigma] (v12) -- (v13);
\draw[phi] (v2) -- (v8);
\draw[phi] (v4) -- (v10);
\draw[fermionnoarrow] (v1) -- (v5);
\draw[fermionnoarrow] (v7) -- (v11);
\end{tikzpicture}
}
\parbox{15mm}{
\begin{tikzpicture}[line width=1 pt,node distance=0.333 cm and 0.3125 cm]
\coordinate[] (v1);
\coordinate[right = of v1] (v2);
\coordinate[right = of v2] (v3);
\coordinate[right = of v4] (v5);
\coordinate[below right = of v2] (v6);
\coordinate[below = of v6] (v7);
\coordinate[below left = of v7] (v8);
\coordinate[below right = of v7] (v9);
\coordinate[left = of v8] (v10);
\coordinate[right = of v9] (v11);
\draw[sigma] (v6) -- (v7);
\draw[phi] (v2) -- (v6);
\draw[phi] (v4) -- (v6);
\draw[phi] (v8) -- (v7);
\draw[phi] (v9) -- (v7);
\draw[fermionnoarrow] (v1) -- (v5);
\draw[fermionnoarrow] (v10) -- (v11);
\end{tikzpicture}
}
\parbox{15mm}{
\begin{tikzpicture}[line width=1 pt,node distance=0.5 cm and 0.25 cm]
\coordinate[] (v1);
\coordinate[right = of v1] (v2);
\coordinate[below = of v2] (v3);
\coordinate[below = of v3] (v4);
\coordinate[right = of v3] (v5);
\coordinate[right = of v5] (v6);
\coordinate[below left = of v6] (v7);
\coordinate[below right = of v6] (v8);
\coordinate[right = of v8] (v9);
\coordinate[left = of v4] (v10);
\coordinate[above right= of v6] (v11);
\coordinate[right= of v11] (v12);
\draw[phi] (v2) -- (v4);
\draw[phi] (v7) -- (v6);
\draw[phi] (v8) -- (v6);
\draw[sigma] (v3) -- (v6);
\draw[fermionnoarrow] (v9) -- (v10);
\draw[fermionnoarrow] (v1) -- (v12);
\end{tikzpicture}
} \right)^2\ .
\end{equation}
This term contains 15 5PN graphs, 5 of which are manifestly vanishing,
while the contribution of the remaining 10 diagrams reads:
\begin{eqnarray}
 {\cal V}_{({\rm 2PN})^2} &=&
\frac{1}{18}\frac{G_N^6 m_1^6 m_2}{r^6}+\frac{11}{6}\frac{G_N^6
                        m_1^5 m_2^2}{r^6} 
\nonumber \\
& &
+\frac{37}{3}\frac{G_N^6 m_1^4 m_2^3}{r^6}+(m_1\leftrightarrow m_2) 
\  . \qquad
\end{eqnarray}

\paragraph{Total 5PN static potential.}
By combining all the previous results, the expression for
the static sector of the 5PN potential finally reads,

\begin{eqnarray}
\label{statictot}
 {\cal V}^{({\rm 5PN})}_{\rm static} &=&  
{\cal V}_{N^6} + {\cal V}_{N^3 \times {\rm 2PN}} + {\cal V}_{N \times {\rm 4PN}} 
+ {\cal V}_{({\rm 2PN})^2}
\nonumber \\
&=&
\frac{5}{16}\frac{G_N^6 m_1^6 m_2}{r^6}+\frac{91}{6}\frac{G_N^6
    m_1^5 m_2^2}{r^6} 
\nonumber \\
& &
+\frac{653}{6}\frac{G_N^6 m_1^4 m_2^3}{r^6}+(m_1\leftrightarrow m_2) \  .
 \qquad
\end{eqnarray}

%
This expression contains the genuine $G_N^6$ contribution coming from
graphs, without contributions generated from lower-$G_N$ terms when using the equations of motion to eliminate terms quadratic at least in the accelerations.
Together with the factorization theorem, the above expressions constitute the second important result of this Letter \footnote{Eq.(\ref{statictot}) has been later confirmed in \cite{Blumlein:2019zku}.}.

\begin{figure}
\begin{center}
\includegraphics[width=1\linewidth]{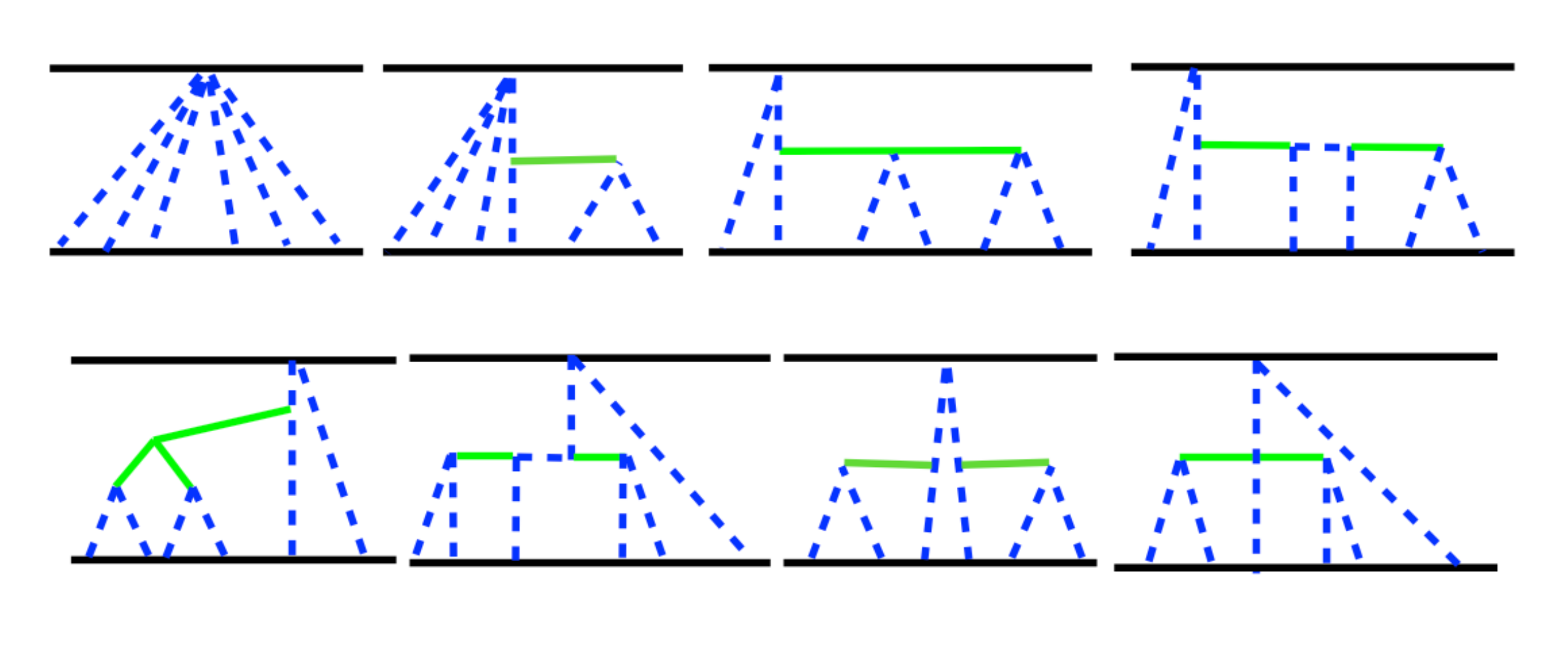}
\caption{5PN graphs contributing to the test-particle limit. The last graph (bottom-right) does not contribute
  to the 5PN potential, because its 4PN subdiagram vanishes.}
\label{fig:test}
\end{center}
\end{figure}

\paragraph{Check: test-particle limit.}
It is possible to verify that the coefficient of the term $m_1^6 m_2$
agrees with what can be expected from the extreme mass ratio limit $m_2\ll m_1$. 
In this limit, where only the graphs displayed in fig.~\ref{fig:test}
contribute, it is possible to consider the body with mass $m_2$ as a test particle in the Schwarzschild metric generated by the body with mass $m_1$.

The action describing the dynamics of the test body has still the form $S_{\rm pp}$ described in eq.(\ref{az_pp}), but with $g_{\mu\nu}$
given by the Schwarzschild metric in harmonic coordinates (which is obtained from the traditional form by the simple radial coordinate shift $r\rightarrow r+  G_N m_1$) instead of the Minkowski one.

In the static limit, $v_2=0$, only the term $g_{00}$ survives, and the effective Lagrangian reads
\be
{\cal L}^{m_2\ll m_1}_{\rm static}=-m_2\sqrt{-g_{00}}=\ds
-m_2\sqrt{\frac{1-\frac{G_N m_1}{r}}{1+\frac{G_N m_1}{r}}} \ .
\ee
By expanding this expression in $\frac{G_N m_1}{r}$, one obtains the sequence\\
$(1,-\frac12,\frac12,-\frac38, \frac38,-\frac{5}{16},\frac5{16},-\frac{35}{128},\frac{35}{128},-\frac{63}{256},\dots)$ of all the coefficients of the $n$PN static terms $G_N^n m_1^n m_2/r^n$, including the $-\frac{5}{16}$ of the 5PN term reported in eq.(\ref{statictot}) (where the potential is correctly  reported with opposite sign with respect to
the lagrangian term).

\section{Conclusion}
\label{sec:conclusion}
We studied the two-body conservative dynamics at fifth
post-Newtonian order (5PN) in the static limit
within the effective field theory approach to General Relativity.
We determined an essential contribution of the complete 5PN potential
at ${\cal O}(G_N^6)$, coming from 154 Feynman diagrams. 
We proved a factorization property of the static diagrams at odd-PN
order, and exploited it to show that their contribution can be
determined recursively, from lower PN orders. 
The result of the static potential at order $G_N^6$ is found to be finite and
rational - a property clearly inherited from the static  $G_N^5$ sector -  
and exhibits the expected Schwarzschild-like behavior in the extreme mass ratio limit. 
The factorization theorem can be applied as well to even-PN
 orders, where it simplifies the evaluation of a large subset of the
 contributing diagrams, therefore 
becoming a powerful tool to systematize and to ease the computations at high-PN
orders.


\acknowledgments
\paragraph{Acknowledgments.}
S.F. is supported by the Fonds National Suisse and by the SwissMap NCCR.
P.M. is supported by the Supporting TAlent in ReSearch at Padova University (UniPD STARS Grant 2017 ``Diagrammalgebra'').
RS is partially supported by CNPq.
W.J.T. has been supported in part by Grants No. FPA2017-84445-P and No. SEV-2014-0398 (AEI/ERDF, EU),
the COST Action CA16201 PARTICLEFACE, and the ``Juan de la Cierva Formaci\'on'' program (FJCI-2017-32128).


\bibliography{references.bib}

\begin{thebibliography}{47}%
\makeatletter
\providecommand \@ifxundefined [1]{%
 \@ifx{#1\undefined}
}%
\providecommand \@ifnum [1]{%
 \ifnum #1\expandafter \@firstoftwo
 \else \expandafter \@secondoftwo
 \fi
}%
\providecommand \@ifx [1]{%
 \ifx #1\expandafter \@firstoftwo
 \else \expandafter \@secondoftwo
 \fi
}%
\providecommand \natexlab [1]{#1}%
\providecommand \enquote  [1]{``#1''}%
\providecommand \bibnamefont  [1]{#1}%
\providecommand \bibfnamefont [1]{#1}%
\providecommand \citenamefont [1]{#1}%
\providecommand \href@noop [0]{\@secondoftwo}%
\providecommand \href [0]{\begingroup \@sanitize@url \@href}%
\providecommand \@href[1]{\@@startlink{#1}\@@href}%
\providecommand \@@href[1]{\endgroup#1\@@endlink}%
\providecommand \@sanitize@url [0]{\catcode `\\12\catcode `\$12\catcode
  `\&12\catcode `\#12\catcode `\^12\catcode `\_12\catcode `\%12\relax}%
\providecommand \@@startlink[1]{}%
\providecommand \@@endlink[0]{}%
\providecommand \url  [0]{\begingroup\@sanitize@url \@url }%
\providecommand \@url [1]{\endgroup\@href {#1}{\urlprefix }}%
\providecommand \urlprefix  [0]{URL }%
\providecommand \Eprint [0]{\href }%
\providecommand \doibase [0]{http://dx.doi.org/}%
\providecommand \selectlanguage [0]{\@gobble}%
\providecommand \bibinfo  [0]{\@secondoftwo}%
\providecommand \bibfield  [0]{\@secondoftwo}%
\providecommand \translation [1]{[#1]}%
\providecommand \BibitemOpen [0]{}%
\providecommand \bibitemStop [0]{}%
\providecommand \bibitemNoStop [0]{.\EOS\space}%
\providecommand \EOS [0]{\spacefactor3000\relax}%
\providecommand \BibitemShut  [1]{\csname bibitem#1\endcsname}%
\let\auto@bib@innerbib\@empty
\bibitem [{\citenamefont {Aasi}\ \emph {et~al.}(2015)\citenamefont {Aasi} \emph
  {et~al.}}]{TheLIGOScientific:2014jea}%
  \BibitemOpen
  \bibfield  {author} {\bibinfo {author} {\bibfnamefont {J.}~\bibnamefont
  {Aasi}} \emph {et~al.} (\bibinfo {collaboration} {LIGO Scientific}),\ }\href
  {\doibase 10.1088/0264-9381/32/7/074001} {\bibfield  {journal} {\bibinfo
  {journal} {Class. Quant. Grav.}\ }\textbf {\bibinfo {volume} {32}},\ \bibinfo
  {pages} {074001} (\bibinfo {year} {2015})},\ \Eprint
  {http://arxiv.org/abs/1411.4547} {arXiv:1411.4547 [gr-qc]} \BibitemShut
  {NoStop}%
\bibitem [{\citenamefont {Acernese}\ \emph {et~al.}(2015)\citenamefont
  {Acernese} \emph {et~al.}}]{TheVirgo:2014hva}%
  \BibitemOpen
  \bibfield  {author} {\bibinfo {author} {\bibfnamefont {F.}~\bibnamefont
  {Acernese}} \emph {et~al.} (\bibinfo {collaboration} {VIRGO}),\ }\href
  {\doibase 10.1088/0264-9381/32/2/024001} {\bibfield  {journal} {\bibinfo
  {journal} {Class. Quant. Grav.}\ }\textbf {\bibinfo {volume} {32}},\ \bibinfo
  {pages} {024001} (\bibinfo {year} {2015})},\ \Eprint
  {http://arxiv.org/abs/1408.3978} {arXiv:1408.3978 [gr-qc]} \BibitemShut
  {NoStop}%
\bibitem [{\citenamefont {Taracchini}\ \emph {et~al.}(2012)\citenamefont
  {Taracchini}, \citenamefont {Pan}, \citenamefont {Buonanno}, \citenamefont
  {Barausse}, \citenamefont {Boyle}, \citenamefont {Chu}, \citenamefont
  {Lovelace}, \citenamefont {Pfeiffer},\ and\ \citenamefont
  {Scheel}}]{Taracchini:2012ig}%
  \BibitemOpen
  \bibfield  {author} {\bibinfo {author} {\bibfnamefont {A.}~\bibnamefont
  {Taracchini}}, \bibinfo {author} {\bibfnamefont {Y.}~\bibnamefont {Pan}},
  \bibinfo {author} {\bibfnamefont {A.}~\bibnamefont {Buonanno}}, \bibinfo
  {author} {\bibfnamefont {E.}~\bibnamefont {Barausse}}, \bibinfo {author}
  {\bibfnamefont {M.}~\bibnamefont {Boyle}}, \bibinfo {author} {\bibfnamefont
  {T.}~\bibnamefont {Chu}}, \bibinfo {author} {\bibfnamefont {G.}~\bibnamefont
  {Lovelace}}, \bibinfo {author} {\bibfnamefont {H.~P.}\ \bibnamefont
  {Pfeiffer}}, \ and\ \bibinfo {author} {\bibfnamefont {M.~A.}\ \bibnamefont
  {Scheel}},\ }\href {\doibase 10.1103/PhysRevD.86.024011} {\bibfield
  {journal} {\bibinfo  {journal} {Phys. Rev.}\ }\textbf {\bibinfo {volume}
  {D86}},\ \bibinfo {pages} {024011} (\bibinfo {year} {2012})},\ \Eprint
  {http://arxiv.org/abs/1202.0790} {arXiv:1202.0790 [gr-qc]} \BibitemShut
  {NoStop}%
\bibitem [{\citenamefont {Schmidt}\ \emph {et~al.}(2015)\citenamefont
  {Schmidt}, \citenamefont {Ohme},\ and\ \citenamefont
  {Hannam}}]{Schmidt:2014iyl}%
  \BibitemOpen
  \bibfield  {author} {\bibinfo {author} {\bibfnamefont {P.}~\bibnamefont
  {Schmidt}}, \bibinfo {author} {\bibfnamefont {F.}~\bibnamefont {Ohme}}, \
  and\ \bibinfo {author} {\bibfnamefont {M.}~\bibnamefont {Hannam}},\ }\href
  {\doibase 10.1103/PhysRevD.91.024043} {\bibfield  {journal} {\bibinfo
  {journal} {Phys. Rev.}\ }\textbf {\bibinfo {volume} {D91}},\ \bibinfo {pages}
  {024043} (\bibinfo {year} {2015})},\ \Eprint {http://arxiv.org/abs/1408.1810}
  {arXiv:1408.1810 [gr-qc]} \BibitemShut {NoStop}%
\bibitem [{\citenamefont {Punturo}\ \emph {et~al.}(2010)\citenamefont {Punturo}
  \emph {et~al.}}]{Punturo:2010zz}%
  \BibitemOpen
  \bibfield  {author} {\bibinfo {author} {\bibfnamefont {M.}~\bibnamefont
  {Punturo}} \emph {et~al.},\ }\bibfield  {booktitle} {\emph {\bibinfo
  {booktitle} {{Proceedings, 14th Workshop on Gravitational wave data analysis
  (GWDAW-14): Rome, Italy, January 26-29, 2010}}},\ }\href {\doibase
  10.1088/0264-9381/27/19/194002} {\bibfield  {journal} {\bibinfo  {journal}
  {Class. Quant. Grav.}\ }\textbf {\bibinfo {volume} {27}},\ \bibinfo {pages}
  {194002} (\bibinfo {year} {2010})}\BibitemShut {NoStop}%
\bibitem [{\citenamefont {Audley}\ \emph {et~al.}(2017)\citenamefont {Audley}
  \emph {et~al.}}]{Audley:2017drz}%
  \BibitemOpen
  \bibfield  {author} {\bibinfo {author} {\bibfnamefont {H.}~\bibnamefont
  {Audley}} \emph {et~al.} (\bibinfo {collaboration} {LISA}),\ }\href@noop {}
  {\  (\bibinfo {year} {2017})},\ \Eprint {http://arxiv.org/abs/1702.00786}
  {arXiv:1702.00786 [astro-ph.IM]} \BibitemShut {NoStop}%
\bibitem [{\citenamefont {Lindblom}\ \emph {et~al.}(2008)\citenamefont
  {Lindblom}, \citenamefont {Owen},\ and\ \citenamefont
  {Brown}}]{Lindblom:2008cm}%
  \BibitemOpen
  \bibfield  {author} {\bibinfo {author} {\bibfnamefont {L.}~\bibnamefont
  {Lindblom}}, \bibinfo {author} {\bibfnamefont {B.~J.}\ \bibnamefont {Owen}},
  \ and\ \bibinfo {author} {\bibfnamefont {D.~A.}\ \bibnamefont {Brown}},\
  }\href {\doibase 10.1103/PhysRevD.78.124020} {\bibfield  {journal} {\bibinfo
  {journal} {Phys. Rev.}\ }\textbf {\bibinfo {volume} {D78}},\ \bibinfo {pages}
  {124020} (\bibinfo {year} {2008})},\ \Eprint {http://arxiv.org/abs/0809.3844}
  {arXiv:0809.3844 [gr-qc]} \BibitemShut {NoStop}%
\bibitem [{\citenamefont {Antonelli}\ \emph {et~al.}(2019)\citenamefont
  {Antonelli}, \citenamefont {Buonanno}, \citenamefont {Steinhoff},
  \citenamefont {van~de Meent},\ and\ \citenamefont
  {Vines}}]{Antonelli:2019ytb}%
  \BibitemOpen
  \bibfield  {author} {\bibinfo {author} {\bibfnamefont {A.}~\bibnamefont
  {Antonelli}}, \bibinfo {author} {\bibfnamefont {A.}~\bibnamefont {Buonanno}},
  \bibinfo {author} {\bibfnamefont {J.}~\bibnamefont {Steinhoff}}, \bibinfo
  {author} {\bibfnamefont {M.}~\bibnamefont {van~de Meent}}, \ and\ \bibinfo
  {author} {\bibfnamefont {J.}~\bibnamefont {Vines}},\ }\href {\doibase
  10.1103/PhysRevD.99.104004} {\bibfield  {journal} {\bibinfo  {journal} {Phys.
  Rev.}\ }\textbf {\bibinfo {volume} {D99}},\ \bibinfo {pages} {104004}
  (\bibinfo {year} {2019})},\ \Eprint {http://arxiv.org/abs/1901.07102}
  {arXiv:1901.07102 [gr-qc]} \BibitemShut {NoStop}%
\bibitem [{\citenamefont {Einstein}\ \emph {et~al.}(1938)\citenamefont
  {Einstein}, \citenamefont {Infeld},\ and\ \citenamefont
  {Hoffmann}}]{Einstein:1938yz}%
  \BibitemOpen
  \bibfield  {author} {\bibinfo {author} {\bibfnamefont {A.}~\bibnamefont
  {Einstein}}, \bibinfo {author} {\bibfnamefont {L.}~\bibnamefont {Infeld}}, \
  and\ \bibinfo {author} {\bibfnamefont {B.}~\bibnamefont {Hoffmann}},\ }\href
  {\doibase 10.2307/1968714} {\bibfield  {journal} {\bibinfo  {journal} {Annals
  Math.}\ }\textbf {\bibinfo {volume} {39}},\ \bibinfo {pages} {65} (\bibinfo
  {year} {1938})}\BibitemShut {NoStop}%
\bibitem [{\citenamefont {Damour}(1982)}]{Damour:1982wm}%
  \BibitemOpen
  \bibfield  {author} {\bibinfo {author} {\bibfnamefont {T.}~\bibnamefont
  {Damour}},\ }in\ \href@noop {} {\emph {\bibinfo {booktitle} {{Les Houches
  Summer School on Gravitational Radiation Les Houches, France, June 2-21,
  1982}}}}\ (\bibinfo {year} {1982})\BibitemShut {NoStop}%
\bibitem [{\citenamefont {Damour}\ and\ \citenamefont
  {Sch{\"a}fer}(1985)}]{Damour:1985mt}%
  \BibitemOpen
  \bibfield  {author} {\bibinfo {author} {\bibfnamefont {T.}~\bibnamefont
  {Damour}}\ and\ \bibinfo {author} {\bibfnamefont {G.}~\bibnamefont
  {Sch{\"a}fer}},\ }\href {\doibase 10.1007/BF00773685} {\bibfield  {journal}
  {\bibinfo  {journal} {Gen. Rel. Grav.}\ }\textbf {\bibinfo {volume} {17}},\
  \bibinfo {pages} {879} (\bibinfo {year} {1985})}\BibitemShut {NoStop}%
\bibitem [{\citenamefont {Damour}\ \emph {et~al.}(2001)\citenamefont {Damour},
  \citenamefont {Jaranowski},\ and\ \citenamefont
  {Sch{\"a}fer}}]{Damour:2001bu}%
  \BibitemOpen
  \bibfield  {author} {\bibinfo {author} {\bibfnamefont {T.}~\bibnamefont
  {Damour}}, \bibinfo {author} {\bibfnamefont {P.}~\bibnamefont {Jaranowski}},
  \ and\ \bibinfo {author} {\bibfnamefont {G.}~\bibnamefont {Sch{\"a}fer}},\
  }\href {\doibase 10.1016/S0370-2693(01)00642-6} {\bibfield  {journal}
  {\bibinfo  {journal} {Phys. Lett.}\ }\textbf {\bibinfo {volume} {B513}},\
  \bibinfo {pages} {147} (\bibinfo {year} {2001})},\ \Eprint
  {http://arxiv.org/abs/gr-qc/0105038} {arXiv:gr-qc/0105038 [gr-qc]}
  \BibitemShut {NoStop}%
\bibitem [{\citenamefont {Blanchet}\ \emph {et~al.}(2004)\citenamefont
  {Blanchet}, \citenamefont {Damour},\ and\ \citenamefont
  {Esposito-Farese}}]{Blanchet:2003gy}%
  \BibitemOpen
  \bibfield  {author} {\bibinfo {author} {\bibfnamefont {L.}~\bibnamefont
  {Blanchet}}, \bibinfo {author} {\bibfnamefont {T.}~\bibnamefont {Damour}}, \
  and\ \bibinfo {author} {\bibfnamefont {G.}~\bibnamefont {Esposito-Farese}},\
  }\href {\doibase 10.1103/PhysRevD.69.124007} {\bibfield  {journal} {\bibinfo
  {journal} {Phys. Rev.}\ }\textbf {\bibinfo {volume} {D69}},\ \bibinfo {pages}
  {124007} (\bibinfo {year} {2004})},\ \Eprint
  {http://arxiv.org/abs/gr-qc/0311052} {arXiv:gr-qc/0311052 [gr-qc]}
  \BibitemShut {NoStop}%
\bibitem [{\citenamefont {Itoh}\ and\ \citenamefont
  {Futamase}(2003)}]{Itoh:2003fy}%
  \BibitemOpen
  \bibfield  {author} {\bibinfo {author} {\bibfnamefont {Y.}~\bibnamefont
  {Itoh}}\ and\ \bibinfo {author} {\bibfnamefont {T.}~\bibnamefont
  {Futamase}},\ }\href {\doibase 10.1103/PhysRevD.68.121501} {\bibfield
  {journal} {\bibinfo  {journal} {Phys. Rev.}\ }\textbf {\bibinfo {volume}
  {D68}},\ \bibinfo {pages} {121501} (\bibinfo {year} {2003})},\ \Eprint
  {http://arxiv.org/abs/gr-qc/0310028} {arXiv:gr-qc/0310028 [gr-qc]}
  \BibitemShut {NoStop}%
\bibitem [{\citenamefont {Damour}\ \emph {et~al.}(2014)\citenamefont {Damour},
  \citenamefont {Jaranowski},\ and\ \citenamefont
  {Sch{\"a}fer}}]{Damour:2014jta}%
  \BibitemOpen
  \bibfield  {author} {\bibinfo {author} {\bibfnamefont {T.}~\bibnamefont
  {Damour}}, \bibinfo {author} {\bibfnamefont {P.}~\bibnamefont {Jaranowski}},
  \ and\ \bibinfo {author} {\bibfnamefont {G.}~\bibnamefont {Sch{\"a}fer}},\
  }\href {\doibase 10.1103/PhysRevD.89.064058} {\bibfield  {journal} {\bibinfo
  {journal} {Phys. Rev.}\ }\textbf {\bibinfo {volume} {D89}},\ \bibinfo {pages}
  {064058} (\bibinfo {year} {2014})},\ \Eprint {http://arxiv.org/abs/1401.4548}
  {arXiv:1401.4548 [gr-qc]} \BibitemShut {NoStop}%
\bibitem [{\citenamefont {Damour}\ \emph {et~al.}(2015)\citenamefont {Damour},
  \citenamefont {Jaranowski},\ and\ \citenamefont
  {Sch{\"a}fer}}]{Damour:2015isa}%
  \BibitemOpen
  \bibfield  {author} {\bibinfo {author} {\bibfnamefont {T.}~\bibnamefont
  {Damour}}, \bibinfo {author} {\bibfnamefont {P.}~\bibnamefont {Jaranowski}},
  \ and\ \bibinfo {author} {\bibfnamefont {G.}~\bibnamefont {Sch{\"a}fer}},\
  }\href {\doibase 10.1103/PhysRevD.91.084024} {\bibfield  {journal} {\bibinfo
  {journal} {Phys. Rev.}\ }\textbf {\bibinfo {volume} {D91}},\ \bibinfo {pages}
  {084024} (\bibinfo {year} {2015})},\ \Eprint
  {http://arxiv.org/abs/1502.07245} {arXiv:1502.07245 [gr-qc]} \BibitemShut
  {NoStop}%
\bibitem [{\citenamefont {Damour}\ \emph {et~al.}(2016)\citenamefont {Damour},
  \citenamefont {Jaranowski},\ and\ \citenamefont
  {Sch{\"a}fer}}]{Damour:2016abl}%
  \BibitemOpen
  \bibfield  {author} {\bibinfo {author} {\bibfnamefont {T.}~\bibnamefont
  {Damour}}, \bibinfo {author} {\bibfnamefont {P.}~\bibnamefont {Jaranowski}},
  \ and\ \bibinfo {author} {\bibfnamefont {G.}~\bibnamefont {Sch{\"a}fer}},\
  }\href {\doibase 10.1103/PhysRevD.93.084014} {\bibfield  {journal} {\bibinfo
  {journal} {Phys. Rev.}\ }\textbf {\bibinfo {volume} {D93}},\ \bibinfo {pages}
  {084014} (\bibinfo {year} {2016})},\ \Eprint
  {http://arxiv.org/abs/1601.01283} {arXiv:1601.01283 [gr-qc]} \BibitemShut
  {NoStop}%
\bibitem [{\citenamefont {Bernard}\ \emph {et~al.}(2016)\citenamefont
  {Bernard}, \citenamefont {Blanchet}, \citenamefont {Boh{\'e}}, \citenamefont
  {Faye},\ and\ \citenamefont {Marsat}}]{Bernard:2015njp}%
  \BibitemOpen
  \bibfield  {author} {\bibinfo {author} {\bibfnamefont {L.}~\bibnamefont
  {Bernard}}, \bibinfo {author} {\bibfnamefont {L.}~\bibnamefont {Blanchet}},
  \bibinfo {author} {\bibfnamefont {A.}~\bibnamefont {Boh{\'e}}}, \bibinfo
  {author} {\bibfnamefont {G.}~\bibnamefont {Faye}}, \ and\ \bibinfo {author}
  {\bibfnamefont {S.}~\bibnamefont {Marsat}},\ }\href {\doibase
  10.1103/PhysRevD.93.084037} {\bibfield  {journal} {\bibinfo  {journal} {Phys.
  Rev.}\ }\textbf {\bibinfo {volume} {D93}},\ \bibinfo {pages} {084037}
  (\bibinfo {year} {2016})},\ \Eprint {http://arxiv.org/abs/1512.02876}
  {arXiv:1512.02876 [gr-qc]} \BibitemShut {NoStop}%
\bibitem [{\citenamefont {Bernard}\ \emph
  {et~al.}(2017{\natexlab{a}})\citenamefont {Bernard}, \citenamefont
  {Blanchet}, \citenamefont {Bohé}, \citenamefont {Faye},\ and\ \citenamefont
  {Marsat}}]{Bernard:2016wrg}%
  \BibitemOpen
  \bibfield  {author} {\bibinfo {author} {\bibfnamefont {L.}~\bibnamefont
  {Bernard}}, \bibinfo {author} {\bibfnamefont {L.}~\bibnamefont {Blanchet}},
  \bibinfo {author} {\bibfnamefont {A.}~\bibnamefont {Bohé}}, \bibinfo
  {author} {\bibfnamefont {G.}~\bibnamefont {Faye}}, \ and\ \bibinfo {author}
  {\bibfnamefont {S.}~\bibnamefont {Marsat}},\ }\href {\doibase
  10.1103/PhysRevD.95.044026} {\bibfield  {journal} {\bibinfo  {journal} {Phys.
  Rev.}\ }\textbf {\bibinfo {volume} {D95}},\ \bibinfo {pages} {044026}
  (\bibinfo {year} {2017}{\natexlab{a}})},\ \Eprint
  {http://arxiv.org/abs/1610.07934} {arXiv:1610.07934 [gr-qc]} \BibitemShut
  {NoStop}%
\bibitem [{\citenamefont {Bernard}\ \emph
  {et~al.}(2017{\natexlab{b}})\citenamefont {Bernard}, \citenamefont
  {Blanchet}, \citenamefont {BohŽ}, \citenamefont {Faye},\ and\ \citenamefont
  {Marsat}}]{Bernard:2017bvn}%
  \BibitemOpen
  \bibfield  {author} {\bibinfo {author} {\bibfnamefont {L.}~\bibnamefont
  {Bernard}}, \bibinfo {author} {\bibfnamefont {L.}~\bibnamefont {Blanchet}},
  \bibinfo {author} {\bibfnamefont {A.}~\bibnamefont {BohŽ}}, \bibinfo {author}
  {\bibfnamefont {G.}~\bibnamefont {Faye}}, \ and\ \bibinfo {author}
  {\bibfnamefont {S.}~\bibnamefont {Marsat}},\ }\href {\doibase
  10.1103/PhysRevD.96.104043} {\bibfield  {journal} {\bibinfo  {journal} {Phys.
  Rev.}\ }\textbf {\bibinfo {volume} {D96}},\ \bibinfo {pages} {104043}
  (\bibinfo {year} {2017}{\natexlab{b}})},\ \Eprint
  {http://arxiv.org/abs/1706.08480} {arXiv:1706.08480 [gr-qc]} \BibitemShut
  {NoStop}%
\bibitem [{\citenamefont {Marchand}\ \emph {et~al.}(2018)\citenamefont
  {Marchand}, \citenamefont {Bernard}, \citenamefont {Blanchet},\ and\
  \citenamefont {Faye}}]{Marchand:2017pir}%
  \BibitemOpen
  \bibfield  {author} {\bibinfo {author} {\bibfnamefont {T.}~\bibnamefont
  {Marchand}}, \bibinfo {author} {\bibfnamefont {L.}~\bibnamefont {Bernard}},
  \bibinfo {author} {\bibfnamefont {L.}~\bibnamefont {Blanchet}}, \ and\
  \bibinfo {author} {\bibfnamefont {G.}~\bibnamefont {Faye}},\ }\href {\doibase
  10.1103/PhysRevD.97.044023} {\bibfield  {journal} {\bibinfo  {journal} {Phys.
  Rev.}\ }\textbf {\bibinfo {volume} {D97}},\ \bibinfo {pages} {044023}
  (\bibinfo {year} {2018})},\ \Eprint {http://arxiv.org/abs/1707.09289}
  {arXiv:1707.09289 [gr-qc]} \BibitemShut {NoStop}%
\bibitem [{\citenamefont {Bernard}\ \emph {et~al.}(2018)\citenamefont
  {Bernard}, \citenamefont {Blanchet}, \citenamefont {Faye},\ and\
  \citenamefont {Marchand}}]{Bernard:2017ktp}%
  \BibitemOpen
  \bibfield  {author} {\bibinfo {author} {\bibfnamefont {L.}~\bibnamefont
  {Bernard}}, \bibinfo {author} {\bibfnamefont {L.}~\bibnamefont {Blanchet}},
  \bibinfo {author} {\bibfnamefont {G.}~\bibnamefont {Faye}}, \ and\ \bibinfo
  {author} {\bibfnamefont {T.}~\bibnamefont {Marchand}},\ }\href {\doibase
  10.1103/PhysRevD.97.044037} {\bibfield  {journal} {\bibinfo  {journal} {Phys.
  Rev.}\ }\textbf {\bibinfo {volume} {D97}},\ \bibinfo {pages} {044037}
  (\bibinfo {year} {2018})},\ \Eprint {http://arxiv.org/abs/1711.00283}
  {arXiv:1711.00283 [gr-qc]} \BibitemShut {NoStop}%
\bibitem [{\citenamefont {Foffa}\ and\ \citenamefont
  {Sturani}(2013)}]{Foffa:2012rn}%
  \BibitemOpen
  \bibfield  {author} {\bibinfo {author} {\bibfnamefont {S.}~\bibnamefont
  {Foffa}}\ and\ \bibinfo {author} {\bibfnamefont {R.}~\bibnamefont
  {Sturani}},\ }\href {\doibase 10.1103/PhysRevD.87.064011} {\bibfield
  {journal} {\bibinfo  {journal} {Phys. Rev.}\ }\textbf {\bibinfo {volume}
  {D87}},\ \bibinfo {pages} {064011} (\bibinfo {year} {2013})},\ \Eprint
  {http://arxiv.org/abs/1206.7087} {arXiv:1206.7087 [gr-qc]} \BibitemShut
  {NoStop}%
\bibitem [{\citenamefont {Foffa}\ \emph {et~al.}(2017)\citenamefont {Foffa},
  \citenamefont {Mastrolia}, \citenamefont {Sturani},\ and\ \citenamefont
  {Sturm}}]{Foffa:2016rgu}%
  \BibitemOpen
  \bibfield  {author} {\bibinfo {author} {\bibfnamefont {S.}~\bibnamefont
  {Foffa}}, \bibinfo {author} {\bibfnamefont {P.}~\bibnamefont {Mastrolia}},
  \bibinfo {author} {\bibfnamefont {R.}~\bibnamefont {Sturani}}, \ and\
  \bibinfo {author} {\bibfnamefont {C.}~\bibnamefont {Sturm}},\ }\href
  {\doibase 10.1103/PhysRevD.95.104009} {\bibfield  {journal} {\bibinfo
  {journal} {Phys. Rev.}\ }\textbf {\bibinfo {volume} {D95}},\ \bibinfo {pages}
  {104009} (\bibinfo {year} {2017})},\ \Eprint
  {http://arxiv.org/abs/1612.00482} {arXiv:1612.00482 [gr-qc]} \BibitemShut
  {NoStop}%
\bibitem [{\citenamefont {Foffa}\ and\ \citenamefont
  {Sturani}(2019)}]{Foffa:2019rdf}%
  \BibitemOpen
  \bibfield  {author} {\bibinfo {author} {\bibfnamefont {S.}~\bibnamefont
  {Foffa}}\ and\ \bibinfo {author} {\bibfnamefont {R.}~\bibnamefont
  {Sturani}},\ }\href@noop {} {\  (\bibinfo {year} {2019})},\ \Eprint
  {http://arxiv.org/abs/1903.05113} {arXiv:1903.05113 [gr-qc]} \BibitemShut
  {NoStop}%
\bibitem [{\citenamefont {Foffa}\ \emph {et~al.}(2019)\citenamefont {Foffa},
  \citenamefont {Porto}, \citenamefont {Rothstein},\ and\ \citenamefont
  {Sturani}}]{Foffa:2019yfl}%
  \BibitemOpen
  \bibfield  {author} {\bibinfo {author} {\bibfnamefont {S.}~\bibnamefont
  {Foffa}}, \bibinfo {author} {\bibfnamefont {R.~A.}\ \bibnamefont {Porto}},
  \bibinfo {author} {\bibfnamefont {I.}~\bibnamefont {Rothstein}}, \ and\
  \bibinfo {author} {\bibfnamefont {R.}~\bibnamefont {Sturani}},\ }\href@noop
  {} {\  (\bibinfo {year} {2019})},\ \Eprint {http://arxiv.org/abs/1903.05118}
  {arXiv:1903.05118 [gr-qc]} \BibitemShut {NoStop}%
\bibitem [{\citenamefont {Ledvinka}\ \emph {et~al.}(2008)\citenamefont
  {Ledvinka}, \citenamefont {Sch{\"a}fer},\ and\ \citenamefont
  {Bicak}}]{Ledvinka:2008tk}%
  \BibitemOpen
  \bibfield  {author} {\bibinfo {author} {\bibfnamefont {T.}~\bibnamefont
  {Ledvinka}}, \bibinfo {author} {\bibfnamefont {G.}~\bibnamefont
  {Sch{\"a}fer}}, \ and\ \bibinfo {author} {\bibfnamefont {J.}~\bibnamefont
  {Bicak}},\ }\href {\doibase 10.1103/PhysRevLett.100.251101} {\bibfield
  {journal} {\bibinfo  {journal} {Phys. Rev. Lett.}\ }\textbf {\bibinfo
  {volume} {100}},\ \bibinfo {pages} {251101} (\bibinfo {year} {2008})},\
  \Eprint {http://arxiv.org/abs/0807.0214} {arXiv:0807.0214 [gr-qc]}
  \BibitemShut {NoStop}%
\bibitem [{\citenamefont {Foffa}(2014)}]{Foffa:2013gja}%
  \BibitemOpen
  \bibfield  {author} {\bibinfo {author} {\bibfnamefont {S.}~\bibnamefont
  {Foffa}},\ }\href {\doibase 10.1103/PhysRevD.89.024019} {\bibfield  {journal}
  {\bibinfo  {journal} {Phys. Rev.}\ }\textbf {\bibinfo {volume} {D89}},\
  \bibinfo {pages} {024019} (\bibinfo {year} {2014})},\ \Eprint
  {http://arxiv.org/abs/1309.3956} {arXiv:1309.3956 [gr-qc]} \BibitemShut
  {NoStop}%
\bibitem [{\citenamefont {Blanchet}\ and\ \citenamefont
  {Fokas}(2018)}]{Blanchet:2018yvb}%
  \BibitemOpen
  \bibfield  {author} {\bibinfo {author} {\bibfnamefont {L.}~\bibnamefont
  {Blanchet}}\ and\ \bibinfo {author} {\bibfnamefont {A.~S.}\ \bibnamefont
  {Fokas}},\ }\href {\doibase 10.1103/PhysRevD.98.084005} {\bibfield  {journal}
  {\bibinfo  {journal} {Phys. Rev.}\ }\textbf {\bibinfo {volume} {D98}},\
  \bibinfo {pages} {084005} (\bibinfo {year} {2018})},\ \Eprint
  {http://arxiv.org/abs/1806.08347} {arXiv:1806.08347 [gr-qc]} \BibitemShut
  {NoStop}%
\bibitem [{\citenamefont {Damour}(2018)}]{Damour:2017zjx}%
  \BibitemOpen
  \bibfield  {author} {\bibinfo {author} {\bibfnamefont {T.}~\bibnamefont
  {Damour}},\ }\href {\doibase 10.1103/PhysRevD.97.044038} {\bibfield
  {journal} {\bibinfo  {journal} {Phys. Rev.}\ }\textbf {\bibinfo {volume}
  {D97}},\ \bibinfo {pages} {044038} (\bibinfo {year} {2018})},\ \Eprint
  {http://arxiv.org/abs/1710.10599} {arXiv:1710.10599 [gr-qc]} \BibitemShut
  {NoStop}%
\bibitem [{\citenamefont {Cheung}\ \emph {et~al.}(2018)\citenamefont {Cheung},
  \citenamefont {Rothstein},\ and\ \citenamefont {Solon}}]{Cheung:2018wkq}%
  \BibitemOpen
  \bibfield  {author} {\bibinfo {author} {\bibfnamefont {C.}~\bibnamefont
  {Cheung}}, \bibinfo {author} {\bibfnamefont {I.~Z.}\ \bibnamefont
  {Rothstein}}, \ and\ \bibinfo {author} {\bibfnamefont {M.~P.}\ \bibnamefont
  {Solon}},\ }\href {\doibase 10.1103/PhysRevLett.121.251101} {\bibfield
  {journal} {\bibinfo  {journal} {Phys. Rev. Lett.}\ }\textbf {\bibinfo
  {volume} {121}},\ \bibinfo {pages} {251101} (\bibinfo {year} {2018})},\
  \Eprint {http://arxiv.org/abs/1808.02489} {arXiv:1808.02489 [hep-th]}
  \BibitemShut {NoStop}%
\bibitem [{\citenamefont {Bern}\ \emph {et~al.}(2019)\citenamefont {Bern},
  \citenamefont {Cheung}, \citenamefont {Roiban}, \citenamefont {Shen},
  \citenamefont {Solon},\ and\ \citenamefont {Zeng}}]{Bern:2019nnu}%
  \BibitemOpen
  \bibfield  {author} {\bibinfo {author} {\bibfnamefont {Z.}~\bibnamefont
  {Bern}}, \bibinfo {author} {\bibfnamefont {C.}~\bibnamefont {Cheung}},
  \bibinfo {author} {\bibfnamefont {R.}~\bibnamefont {Roiban}}, \bibinfo
  {author} {\bibfnamefont {C.-H.}\ \bibnamefont {Shen}}, \bibinfo {author}
  {\bibfnamefont {M.~P.}\ \bibnamefont {Solon}}, \ and\ \bibinfo {author}
  {\bibfnamefont {M.}~\bibnamefont {Zeng}},\ }\href {\doibase
  10.1103/PhysRevLett.122.201603} {\bibfield  {journal} {\bibinfo  {journal}
  {Phys. Rev. Lett.}\ }\textbf {\bibinfo {volume} {122}},\ \bibinfo {pages}
  {201603} (\bibinfo {year} {2019})},\ \Eprint
  {http://arxiv.org/abs/1901.04424} {arXiv:1901.04424 [hep-th]} \BibitemShut
  {NoStop}%
\bibitem [{\citenamefont {Goldberger}\ and\ \citenamefont
  {Rothstein}(2006)}]{Goldberger:2004jt}%
  \BibitemOpen
  \bibfield  {author} {\bibinfo {author} {\bibfnamefont {W.~D.}\ \bibnamefont
  {Goldberger}}\ and\ \bibinfo {author} {\bibfnamefont {I.~Z.}\ \bibnamefont
  {Rothstein}},\ }\href {\doibase 10.1103/PhysRevD.73.104029} {\bibfield
  {journal} {\bibinfo  {journal} {Phys. Rev.}\ }\textbf {\bibinfo {volume}
  {D73}},\ \bibinfo {pages} {104029} (\bibinfo {year} {2006})},\ \Eprint
  {http://arxiv.org/abs/hep-th/0409156} {arXiv:hep-th/0409156 [hep-th]}
  \BibitemShut {NoStop}%
\bibitem [{\citenamefont {Goldberger}(2007)}]{Goldberger:2007hy}%
  \BibitemOpen
  \bibfield  {author} {\bibinfo {author} {\bibfnamefont {W.~D.}\ \bibnamefont
  {Goldberger}},\ }in\ \href@noop {} {\emph {\bibinfo {booktitle} {{Les Houches
  Summer School - Session 86: Particle Physics and Cosmology: The Fabric of
  Spacetime Les Houches, France, July 31-August 25, 2006}}}}\ (\bibinfo {year}
  {2007})\ \Eprint {http://arxiv.org/abs/hep-ph/0701129} {arXiv:hep-ph/0701129
  [hep-ph]} \BibitemShut {NoStop}%
\bibitem [{\citenamefont {Foffa}\ and\ \citenamefont
  {Sturani}(2014)}]{Foffa:2013qca}%
  \BibitemOpen
  \bibfield  {author} {\bibinfo {author} {\bibfnamefont {S.}~\bibnamefont
  {Foffa}}\ and\ \bibinfo {author} {\bibfnamefont {R.}~\bibnamefont
  {Sturani}},\ }\href {\doibase 10.1088/0264-9381/31/4/043001} {\bibfield
  {journal} {\bibinfo  {journal} {Class. Quant. Grav.}\ }\textbf {\bibinfo
  {volume} {31}},\ \bibinfo {pages} {043001} (\bibinfo {year} {2014})},\
  \Eprint {http://arxiv.org/abs/1309.3474} {arXiv:1309.3474 [gr-qc]}
  \BibitemShut {NoStop}%
\bibitem [{\citenamefont {Rothstein}(2014)}]{Rothstein:2014sra}%
  \BibitemOpen
  \bibfield  {author} {\bibinfo {author} {\bibfnamefont {I.~Z.}\ \bibnamefont
  {Rothstein}},\ }\href {\doibase 10.1007/s10714-014-1726-y} {\bibfield
  {journal} {\bibinfo  {journal} {Gen. Rel. Grav.}\ }\textbf {\bibinfo {volume}
  {46}},\ \bibinfo {pages} {1726} (\bibinfo {year} {2014})}\BibitemShut
  {NoStop}%
\bibitem [{\citenamefont {Porto}(2016)}]{Porto:2016pyg}%
  \BibitemOpen
  \bibfield  {author} {\bibinfo {author} {\bibfnamefont {R.~A.}\ \bibnamefont
  {Porto}},\ }\href {\doibase 10.1016/j.physrep.2016.04.003} {\bibfield
  {journal} {\bibinfo  {journal} {Phys. Rept.}\ }\textbf {\bibinfo {volume}
  {633}},\ \bibinfo {pages} {1} (\bibinfo {year} {2016})},\ \Eprint
  {http://arxiv.org/abs/1601.04914} {arXiv:1601.04914 [hep-th]} \BibitemShut
  {NoStop}%
\bibitem [{\citenamefont {Damour}\ and\ \citenamefont
  {Jaranowski}(2017)}]{Damour:2017ced}%
  \BibitemOpen
  \bibfield  {author} {\bibinfo {author} {\bibfnamefont {T.}~\bibnamefont
  {Damour}}\ and\ \bibinfo {author} {\bibfnamefont {P.}~\bibnamefont
  {Jaranowski}},\ }\href {\doibase 10.1103/PhysRevD.95.084005} {\bibfield
  {journal} {\bibinfo  {journal} {Phys. Rev.}\ }\textbf {\bibinfo {volume}
  {D95}},\ \bibinfo {pages} {084005} (\bibinfo {year} {2017})},\ \Eprint
  {http://arxiv.org/abs/1701.02645} {arXiv:1701.02645 [gr-qc]} \BibitemShut
  {NoStop}%
\bibitem [{\citenamefont {Levi}(2018)}]{Levi:2018nxp}%
  \BibitemOpen
  \bibfield  {author} {\bibinfo {author} {\bibfnamefont {M.}~\bibnamefont
  {Levi}},\ }\href@noop {} {\  (\bibinfo {year} {2018})},\ \Eprint
  {http://arxiv.org/abs/1807.01699} {arXiv:1807.01699 [hep-th]} \BibitemShut
  {NoStop}%
\bibitem [{\citenamefont {Foffa}\ and\ \citenamefont
  {Sturani}(2011)}]{Foffa:2011ub}%
  \BibitemOpen
  \bibfield  {author} {\bibinfo {author} {\bibfnamefont {S.}~\bibnamefont
  {Foffa}}\ and\ \bibinfo {author} {\bibfnamefont {R.}~\bibnamefont
  {Sturani}},\ }\href {\doibase 10.1103/PhysRevD.84.044031} {\bibfield
  {journal} {\bibinfo  {journal} {Phys. Rev.}\ }\textbf {\bibinfo {volume}
  {D84}},\ \bibinfo {pages} {044031} (\bibinfo {year} {2011})},\ \Eprint
  {http://arxiv.org/abs/1104.1122} {arXiv:1104.1122 [gr-qc]} \BibitemShut
  {NoStop}%
\bibitem [{\citenamefont {Blanchet}(2014)}]{Blanchet:2014}%
  \BibitemOpen
  \bibfield  {author} {\bibinfo {author} {\bibfnamefont {L.}~\bibnamefont
  {Blanchet}},\ }\href {\doibase 10.12942/lrr-2014-2} {\bibfield  {journal}
  {\bibinfo  {journal} {Living Reviews in Relativity}\ }\textbf {\bibinfo
  {volume} {17}},\ \bibinfo {pages} {2} (\bibinfo {year} {2014})}\BibitemShut
  {NoStop}%
\bibitem [{\citenamefont {Blanchet}\ and\ \citenamefont
  {Damour}(1989)}]{Blanchet:1989ki}%
  \BibitemOpen
  \bibfield  {author} {\bibinfo {author} {\bibfnamefont {L.}~\bibnamefont
  {Blanchet}}\ and\ \bibinfo {author} {\bibfnamefont {T.}~\bibnamefont
  {Damour}},\ }\href@noop {} {\bibfield  {journal} {\bibinfo  {journal} {Ann.
  Inst. H. Poincare Phys. Theor.}\ }\textbf {\bibinfo {volume} {50}},\ \bibinfo
  {pages} {377} (\bibinfo {year} {1989})}\BibitemShut {NoStop}%
\bibitem [{\citenamefont {Kol}\ and\ \citenamefont
  {Smolkin}(2008{\natexlab{a}})}]{Kol:2007bc}%
  \BibitemOpen
  \bibfield  {author} {\bibinfo {author} {\bibfnamefont {B.}~\bibnamefont
  {Kol}}\ and\ \bibinfo {author} {\bibfnamefont {M.}~\bibnamefont {Smolkin}},\
  }\href {\doibase 10.1088/0264-9381/25/14/145011} {\bibfield  {journal}
  {\bibinfo  {journal} {Class. Quant. Grav.}\ }\textbf {\bibinfo {volume}
  {25}},\ \bibinfo {pages} {145011} (\bibinfo {year} {2008}{\natexlab{a}})},\
  \Eprint {http://arxiv.org/abs/0712.4116} {arXiv:0712.4116 [hep-th]}
  \BibitemShut {NoStop}%
\bibitem [{\citenamefont {Kol}\ and\ \citenamefont
  {Smolkin}(2008{\natexlab{b}})}]{Kol:2007rx}%
  \BibitemOpen
  \bibfield  {author} {\bibinfo {author} {\bibfnamefont {B.}~\bibnamefont
  {Kol}}\ and\ \bibinfo {author} {\bibfnamefont {M.}~\bibnamefont {Smolkin}},\
  }\href {\doibase 10.1103/PhysRevD.77.064033} {\bibfield  {journal} {\bibinfo
  {journal} {Phys. Rev.}\ }\textbf {\bibinfo {volume} {D77}},\ \bibinfo {pages}
  {064033} (\bibinfo {year} {2008}{\natexlab{b}})},\ \Eprint
  {http://arxiv.org/abs/0712.2822} {arXiv:0712.2822 [hep-th]} \BibitemShut
  {NoStop}%
\bibitem [{Note1()}]{Note1}%
  \BibitemOpen
  \bibinfo {note} {It is understood that spatial indices in this expression are
  contracted by means of the spatial metric $\gamma _{ij}$, which implies the
  appearance of extra $\sigma $ fields.}\BibitemShut {Stop}%
\bibitem [{Note2()}]{Note2}%
  \BibitemOpen
  \bibinfo {note} {Eq.(\ref {statictot}) has been later confirmed in \cite
  {Blumlein:2019zku}.}\BibitemShut {Stop}%
\bibitem [{\citenamefont {Bl{\"u}mlein}\ \emph {et~al.}(2019)\citenamefont
  {Bl{\"u}mlein}, \citenamefont {Maier},\ and\ \citenamefont
  {Marquard}}]{Blumlein:2019zku}%
  \BibitemOpen
  \bibfield  {author} {\bibinfo {author} {\bibfnamefont {J.}~\bibnamefont
  {Bl{\"u}mlein}}, \bibinfo {author} {\bibfnamefont {A.}~\bibnamefont {Maier}},
  \ and\ \bibinfo {author} {\bibfnamefont {P.}~\bibnamefont {Marquard}},\
  }\href@noop {} {\  (\bibinfo {year} {2019})},\ \Eprint
  {http://arxiv.org/abs/1902.11180} {arXiv:1902.11180 [gr-qc]} \BibitemShut
  {NoStop}%
\end{thebibliography}%

\end{document}